\documentclass[twocolumn,prb,showpacs,amsmath,amssymb]{revtex4}

\usepackage{graphicx}

\begin{document}


\title {\bf 
Condensation, excitation, pairing, and superfluid density in high-$T_{c}$ 
superconductors:
magnetic resonance mode as a roton analogue and 
a possible spin-mediated pairing\/} 
\author{Y.J.~Uemura}
\affiliation{Physics Department, Columbia University, 538 West, 120th Street,
New York, NY 10027, USA}
\date{\today}
\begin{abstract}  
{To find out a primary determing factor of $T_{c}$
and a pairing mechanism
in high-$T_{c}$ cuprates, we combine 
the muon spin relaxation results on $n_{s}/m^{*}$
(superconducting carrier density / effective mass), 
accumulated over the last 15 years, with the results from 
neutron and Raman scattering, 
STM, specific heat, Nernst effect and ARPES measurements.  
We identify the neutron magnetic resonance mode as
an analogue of roton minimum in the superfluid $^{4}$He, and argue
that $n_{s}/m^{*}$ and the resonance mode energy
$\hbar\omega_{res}$ play a primary role in determining $T_{c}$
in the underdoped region.  We propose a picture that 
roton-like excitations in the cuprates appear as a 
coupled mode, which
has the resonance mode for 
spin and charge responses 
at different momentum transfers but the same energy 
transfers, as detected respectively, by 
the neutron S=1 mode and the Raman S=0 A1$_{g}$ mode.
We shall call this as
the ``hybrid spin/charge roton''. 
After discussing the role of dimensionality in 
condensation, we propose a generic phase diagram
of the cuprates with spatial phase separation in 
the overdoped region as a special case of the BE-BCS crossover
conjecture where the superconducting coupling is lost
rapidly in the overdoped region.  Using a microscopic model of
charge motion resonating with 
antiferomagnetic spin fluctuations, we propose
a possibility that the hybrid spin/charge roton
and higher-energy spin fluctuations
mediate the superconducting
pairing.
In this model,  
the resonance modes can be viewed as a meson-analogue
and the ``dome'' shape of the
phase diagram can be understood as 
a natural consequence of departure from the 
competing Mott insulator ground state via carrier
doping.}

\end{abstract}
%

\pacs{74.20.Mn, 74.62.-c, 74.72.-h, 76.75.+i}
\maketitle

\parskip 0cm


\section{INTRODUCTION}

Since the discovery of high-$T_{c}$ superconductors (HTSC), 
muon spin relaxation ($\mu$SR) measurements [1-3]
have been applied to study magnetic order, superconducting penetration 
depth, and 
flux vortex aspects of the cuprates and other exotic superconductors.
Among them, we have put a particular focus
on the absolute values of the magnetic field penetration depth
$\lambda$, which represents the superconducting carrier density $n_{s}$
divided by the effective mass $m^{*}$ as $\lambda^{-2} \propto n_{s}/m^{*}$.
We discovered strong and universal correlations between 
the critical temperature $T_{c}$ and $n_{s}/m^{*}(T\rightarrow 0)$
in various HTSC systems [4], and initiated an energy-scale 
argument by converting
$n_{s}/m^{*}$ into an effective Fermi temperature $T_{F}$ of 
the superconducting 
carriers [5].  By combining the universal correlations 
and the pseudo-gap behavior,
we proposed a picture for the cuprates based on 
crossover from Bose-Einstein (BE)
to BCS condensation in 1994 [6,7].  
We also demonstrated that the correlations between $T_{c}$ and $n_{s}/m^{*}$
are robust against various artificial/spontaneous formations of 
spatial heterogeneity, such as Cu/Zn substitution, for which 
we proposed the ``Swiss Cheese Model'' in 1996 [8].

During recent years after 1997, several important 
experimental results on the cuprates, relevant to the above mentioned
results/pictures, have been obtained by using 
other techniques.  They include:
1) scanning tunnelling microscope results which confirmed the 
``swiss-cheese'' like situation in Zn doped cuprates [9], and further
established spontaneous formation of spatial heterogeneity in 
the superconducting state in a wide range of doping levels even without
impurity substitution [10,11];
2) high-frequency conductivity measurements of $n_{s}/m^{*}$,   
which showed a frequency-dependent ``dynamic'' response 
remaining above $T_{c}$
and provided the first direct evidence for a wide range of 
superconducting phase fluctuations
in the normal state [12];
3) the Nernst effect measurements which obtained further evidence for 
the dynamic superconducting response above $T_{c}$, and mapped out the  
region of this response as a function of external field $H$, temperature $T$,
and the doping concentration $x$ [13,14];
4) neutron scattering studies which established that  
the ``41 meV magnetic resonance mode'', first found in YBCO [15], 
is generic to many of the cuprate systems,
and that the mode energy accurately scales with $T_{c}$ [16-18]; 
5) Raman scattering studies which obtained the response in the 
A1$_{g}$ mode at $T \rightarrow 0$ 
whose energy precisely follows that of the neutron 
resonance peak [19-22]; and
6) a sharp ``coherence peak'' which appears in the antinodal response
of ARPES studies, with the intensity proportional to the superfluid
density [23].

In this paper, we wish to demonstrate that the combination of the
$\mu$SR results with these new results 1) - 6)  
can clarify the primary determining factor for the critical temperature $T_{c}$ 
in the cuprates and can elucidate details of their superconducting
condensation and pairing mechanisms.  This discussion will proceed in 
comparison with known cases of 
quantum superfluid transitions in bulk and thin films of liquid $^{4}$He.
In particular, we will propose a picture that the 41 meV resonance mode
in the cuprates plays a role analogous to rotons in liquid He, as the 
key elementary excitation mode which determines $T_{c}$.
We further propose a picture of a ``hybrid spin/charge roton'' 
in an effort to reconcile apparently 
conflicting selection rules for the
neutron magnetic resonance mode and Raman A1$_{g}$ mode,
which are observed with the same energy transfers [19].

Consideration of dimensionality effects will lead to 
a new and simple account for the onset 
temperature of the 
Nernst effect and conditions necessary for
dynamic superconductivity.  We will compare  
evidence for phase separation in the 
overdoped cuprates, and argue that the condensation in the cuprates can be
viewed as a special case of BE-BCS crossover where
the pairing interaction vanishes in a moderately overdoped region.     
Finally, we consider a microscopic model where the motion of a charge 
is assisted by spin fluctuations having the {\it same resonating frequency\/}.
Developing this notion, we 
propose a picture that 
the ``hybrid spin/charge roton''
and higher-energy antiferromagnetic (AF) spin fluctuations
can be the pair-mediating bosons.
These concepts naturally lead to a view in which we ascribe the anomalies in 
the overdoped region to the rapid disappearence of spin fluctuations  
due to increasing distance from the Mott AF insulator state.

\section{Plots of $T_{c}$ vs. $n_{s}/m^{*}$ and 
$T_{c}$ vs. $T_{F}$}

In the vortex state of type-II superconductors, the applied 
field $H$ enters into 
the system by forming a lattice of flux vortices, with a typical 
distance between the adjacent vortices $\sim 1,000$ \AA\ for $H \sim 1$ kG.
The vortex lattice and supercurrent create inhomogeneity 
$\Delta H$ in the internal magnetic fields, whose decay from the vortex core
position is related to the London penetration depth $\lambda$.  
In $\mu$SR measurements, a time histogram is composed of the 
precession of 10$^{6}$ - 10$^{7}$ muons stopped at
locations of different internal fields.  
In measurements with ceramic specimens, the damping of the precession envelope
is often approximated by a Gaussian function
$\exp(-\sigma^{2}t^{2}/2)$,  which defines the muon spin relaxation rate 
$\sigma \propto \Delta H$.  With the London equation, one
finds 
$$\sigma \propto \lambda^{-2} = 
[4\pi n_{s}e^{2}/m^{*}c^{2}][1/(1+\xi/{\it l\/})],$$
where $\xi$ is the coherence length and {\it l\/} denotes the mean free path.
In systems in the ``clean limit'' with $\xi << {\it l\/}$, such as the cuprates,
$\sigma$ represents $n_{s}/m^{*}$.

Figure 1 shows the accumulation of our results over the last 15 years in a plot of
$T_{c}$ versus $\sigma(T\rightarrow 0) \propto n_{s}/m^{*}$ in 
various cuprate systems [4,5,8,24-27].  These data are consistent
with results from other groups [28-31].  
We see that $T_{c}$ increases with increasing carrier doping, following a
nearly linear relationship with $n_{s}/m^{*}$ in the underdoped region:
the results from different series of cuprates share a common slope in this
initial increase of $T_{c}$.  Systems in the optimally doped region deviate
from this linear line, showing a ``plateau'' like behavior, which was later
found out to be due to the effect of the CuO chains in the case of
the YBCO systems [29].  We see that the nearly linear trend is followed
also by systems with Cu/Zn substitution [8], 
spontaneous formation of static magnetic regions (shown with the ``stripe'' symbols),
and even with overdoping (Tl2201) [24], 
the implications of which will be discussed in later sections.

Strong correlations between $T_{c}$ and $n_{s}/m^{*}$ would not be expected from 
BCS theory [32], but readily expected in BE condensation.  Thus, this figure
gives the first strong message about possible non-BCS character of 
condensation in the cuprates.
In fact, when one assumes $m^{*}$ of the cuprates to be 3-4 times the bare electron mass,
the carrier density $n_{s}$ for systems in the linear region corresponds to the 
situation where several pairs are overlapping within an area of $\pi\xi^{2}$
on the CuO$_{2}$ planes, which interpolates between non-overlapping pairs in the BE limit
and thousands of pairs per $\pi\xi^{2}$ in the BCS limit [33].
Although most of the points in Fig. 1 are obtained with un-oriented ceramic specimens,
the results predominantly reflect the in-plane penetration depth and the in-plane mass,
as generally demonstrated for highly anisotropic superconductors [34] with  
$\sim$ 40 \%\ correction factor between the relaxation rates from 
un-oriented and oriented (or single crystal) 
specimens.

The parameter $n_{s}/m^{*}$ in eq. 1 represents a spectral weight of 
3-dimensional screening supercurrent which causes a partial rejection of the 
applied field.  By knowing the average interlayer distance 
$c_{int}$ between the CuO$_{2}$ planes,
one can obtain the 2-d area density $n_{s2d}$ of superconducting carriers, which 
can be directly converted into the effective Fermi temperature 
$T_{F} \propto n_{s2d}/m^{*} = n_{s}/m^{*} \times c_{int}$.  For 3-d systems,
the results of $\sigma \propto n_{s}/m^{*}$ has to be combined with those
of the Pauli susceptibility or Sommerfeld constant $\gamma \propto n^{1/3}m^{*}$ 
in the derivation of $T_{F} \propto n_{s}^{2/3}/m^{*}$.  After this processing,
we produced a plot of $T_{c}$ versus $T_{F}$ first in 1991 [5].  Figure 2 shows the
up-to-date version of this plot, including subsequent data from varous different
superconductors [33,35-39]

The $T_{B}$ line represents the BE
condensation temperature of an ideal Bose gas of boson mass 2m$^{*}$ and density $n_{s}/2$.
Although the actual superconducting $T_{c}$ of the cuprates are reduced by a factor 4-5
from $T_{B}$, the trend of the underdoped cuprates is parallel to the $T_{B}$ line,
suggesting that the linear trend can be deeply related to BE condensation.
There seems to be an empirical upper limit of $T_{c}/T_{F}$ shared not only by the cuprates
but also by other strongly correlated systems, such as 
organic 2-d BEDT [37], Na$_{x}$CoO$_{2}$ [39], and A$_{3}$C$_{60}$ [35,36] systems.
Figure 2 serves for classifying different superconductors between the limits
of BE condensation with strong coupling (approaching towards the $T_{B}$ line) 
and BCS condensation with much smaller values of $T_{c}/T_{F}$. 

In Fig. 2, we also include a point for superfluid bulk $^{4}$He with $T_{c}$ = 2.2 K
(blue star).  Even for such a system regarded as a prototype for BE condensation,
we see a 30 \%\ reduction from its ideal-gas value of $T_{BE}$ = 3.2 K, presumably
caused by the finite size and interactions of He atoms.  In addition to these factors,
2-dimensional aspects of the cuprates could also cause a further reduction of $T_{c}$ 
from $T_{B}$ as discussed in Section V.  
Recently, condensation of fermionic ultracold $^{40}$K gas has been 
achieved in the BE-BCS crossover region [40].  We also plot a point for $^{40}$K (red star)
by multiplying 10$^{8}$ to both $T_{c}$ and $T_{F}$.  Interestingly, the 
crossover region seems to exist rather close to the $T_{B}$ line for the case of 
ultra-cold fermion atoms in which we expect a much smaller effect of correlations
compared to the cuprates.

\section{BE-BCS crossover conjecture}

A connection between BE and BCS condensation has been a subject of theoretical 
interest for many decades.  There exist a few different energy scales for 
superconductivity: a) the energy of condensing carriers represented by
$T_{F}$, related to the number density and mass; b) the net
attractive interaction, related to the gap energy scale in the BCS side, and 
to the binding energy in local bosons, and c) the energy of ``pair mediating
bosons'' such as the Debye frequency in the phonon-coupling.
The fourth energy scale d), the condensation temperature $T_{c}$
is determined by the interplay among a), b), and c).   
Many theoretical discussions have been given, 
by fixing the carrier concentration (a),
and artificially changing the attractive coupling (b).  As a typical example,
Fig. 3(a) shows the results obtained by Nozi\`eres and Schmitt-Rink [41] in 
1985.  Note that the horizontal 
axis is given for the coupling strength $V$, normalized
to the ``critical strength'' $V_{c}$ related 
to the particle density and the range of the
interaction.  We see that $T_{c} \sim T_{B}$ 
in the strong coupling limit ($1/V < 1$), while $T_{c} \propto V$ 
on the BCS side ($1/V >> 1$).

In cuprate systems, we change the carrier density (a), while 
the attractive energy
scale (b) is not necessarily fixed. On the other hand 
the ``mediating boson'' would not
change for different carrier densities.  So, it would make sense to consider the
case by fixing (c) and varying (a), and consider how (b) and (d) would evolve.  
By combining the results in Figs. 1 and 2 with the pseudogap behavior
which was then-noticed in NMR (spin gap) [42,43] and c-axis conductivity [44] 
(insulating below $T^{*}$) studies, we proposed a picture 
shown in Fig. 3(b) in 1994 [6,7]
to map the situation of the cuprates to BE-BCS crossover.
We argued that the underdoped region corresponds to the BE side,
where the pair formation temperature 
$T_{pair}$ maps to $T^{*}$ below which one 
expects gradual formation of the spin singlet pairs.  On the 
BCS side $T_{pair}$ and $T_{c}$ should
be very close, since the condensation occurs 
immediately after the pair formation.
These two regions may be separated by whether the 
interaction is non-retarded (BE side)
or retarded (BCS) side.  

Independent from our picture, Emery and Kivelson [45] 
presented the phase-fluctuation 
picture shown in Fig. 3(c) in 1995, and interpreted 
the $\mu$SR results as a signature
for a Kosterlitz-Thouless (KT) type transition [46] 
expected in 2-dimensional systems where
superconductivity is destroyed by thermal excitations of phase fluctuations.
Although their picture in Fig. 3(c) shares various fundamental spirits with 
the BE-BCS crossover pictures, some important differences exist as 
discussed in later sections.  The overdoped region of
the cuprates, which corresponds to the ``BCS'' side 
of Figs. 3(b) or 3(c), turns
out to be quite different from the simple BCS systems 
as discussed in Section VIII.  Various types of models based on BE condensation [47,48]
and BE-BCS crossover [49] have also been proposed to account for HTSC.

\section{Heterogeneity}

The relevance of BE condensation also appears in the robustness 
of the cuprates against
heterogenous spatial media.  The first signature of this came in     
our study of Zn-doped YBCO and LSCO systems in 1996 [8].  
As shown in Fig. 4(a), the reduction of 
$n_{s}/m^{*}$, as a function of Zn concentration, 
can be explained very well if we assume
that Zn creates a non-superconducting region 
around it in an area of $\pi\xi^{2}$ on the
CuO$_{2}$ planes.  The solid line in Fig. 4(a), 
representing this ``swiss cheese model''
agrees well with the data {\it without any fitting\/}.  
A few years later, this situation was directly confirmed 
by the scanning tunnelling microscope (STM) studies [9], as shown
in Fig. 4(b).  

In Eu doped (La,Eu,Sr)$_{2}$CuO$_{4}$ (LESCO), regions with static
incommensurate magnetic
order is formed spontaneously even in superconducting specimens [26].  
The volume fraction of 
regions with static magnetic order shows 
a one-to-one trade off with the total value of 
$n_{s}/m^{*}$, as shown in Fig. 4(c), indicating that 
the ``magnetic island'' regions
do not support superfluid.  Thus the situation looks 
again like the ``swiss cheese'',
with ``non-superconducting magnetic islands'' replacing the ``normal area
around Zn''.  
In zero-field
$\mu$SR measurements of oxygen overdoped La$_{2}$CuO$_{4.11}$ [25], we found 
the average size of such magnetic islands to be comparable to the 
in-plane coherence length $\xi\sim$ 15-30 \AA\ in radius.  
The Zn-doped and magnetic-island systems
follow the same trend as other less perturbed underdoped cuprates
in Fig. 1.  

Recently, spontaneous formation of spatially heterogeneous areas, with similar
length scales, was discovered by a series of STM measurements [10,11] in under to 
optimally doped Bi2212 systems without impurity doping.  As shown in Fig. 4(d), 
the STM ``gap map''  clearly demonstrates a decomposition of the system into 
regions with a sharp superconducing gap (orange to red colors) and a broad
pseudo-gap like response (blue to black colors).  Upon doping carriers, 
the former region increases in volume fraction, which roughly
follows the behavior of $n_{s}/m^{*}$ [50].   
The new STM results suggest a possibility that the underdoped 
cuprates are analogous to ``spontaneously formed swiss cheese''.
We shall see later that overdoped Tl2201 also exhibits
yet another type of spatial heterogeneity / phase separation.  

Thus the nearly linear trend in Fig. 1 for all these systems represents
a generic feature of the cuprate superfluid, which is impressively robust 
against the formation of spatial heterogeneity.  This is a feature which 
we can expect for a superfluid of tightly bound bosons, as 
we shall see in the next section, but not for
BCS systems where impurities can scatter an individual carrier
before it appreciates an attractive interaction with 
its pair in a retarded process. 

\section{Two dimensional aspects}

Let us now compare the cuprates with the superfluid transitions in 
thin films of $^{4}$He and $^{4}$He/$^{3}$He
mixtures in regular and porous media.
Figure 5(a) shows the temperature dependence of the superfluid density
of a $^{4}$He film on Mylar [51] and the relaxation rate $\sigma(T)$ in oriented  
optimally-doped YBCO [52]  (multiplied by a correction factor of 1/1.4)
and (Y,Pr) substituted underdoped YBCO [27].  
The superfluid density of $^{4}$He film is almost
independent of temperature until the KT transition 
suddenly drives the system into the normal state.  In contrast, the
superfluid density of YBCO 
shows a strong reduction with increasing $T$: the low-$T$ variation
was attributed to the excitation of nodal quasiparticles in the d-wave
energy gap [53].  

Figure 5(b) shows a plot of the superfluid transition temperature
$T_{c}$ versus the 2-d superfluid density of $^{4}$He films on 
regular (Mylar) [51] and porous (Vycor) [54] media and $^{4}$He/$^{3}$He
mixture adsorbed on fine alumina powders [55].  The Mylar results 
represent the case with the smallest perturbation, comparable to the 
simple hole-doped cuprates.  The Vycor results are analogous
to the Zn-doped cuprates: normal regions are formed
around the pore wall or around Zn impurities in order to 
help smooth flow of the superfluid.  The $^{4}$He/$^{3}$He
mixture corresponds to the overdoped cuprates as discussed
later.  Yet with different degrees of perturbation, 
$T_{c}$ precisely follows the linear relationship predicted by the 
Kosterlitz-Thouless theory [46].  The robustness against heterogeneity
is common to the case of the cuprates in Fig. 1.

By multiplying the interlayer distance $c_{int}$ with $n_{s}/m^{*}$,
we can produce a corresponding plot for the cuprates with the horizontal
axis representing 2-d area superfluid density $n_{s2d}/m^{*}$,
as shown in Fig. 5(c) [56].  We find that: (1) the systems with different 
$c_{int}$ exhibit distinctly different slopes; (2) the 
$T\rightarrow 0$ values of $n_{s2d}/m^{*}$ is more than a factor of 2 
larger than the ``KT-jump'' value $n_{s2d}/m^{*}(T=T_{KT})$ 
expected by the KT theory shown by the $T_{KT}$ line [57]; and (3) 
$T_{c} \propto 1/c_{int}$ from  
the early studies of YBCO/PrBCO sandwich layer systems [58].
These features indicate that the simplest version of the KT 
theory for 2-dimensional
systems is not adequate to account for the transition temperatures of
cuprtates.  This point was missing in the argument of Emery and 
Kivelson [45].  

For a non-interacting Bose gas with a parabollic energy 
dispersion, BE condensation does not occur in purely 
2-dimension.  However, one can expect condensation via 
inclusion of a small 3-d coupling [47,48]
and/or dispersion with the power to $k$ lower
than 2 [59].   A small 3-d coupling for a quasi 2-d Bose gas would
provide a logarithmic dependence of $T_{c}$ on the coupling
strength [47,48], which decays exponentially with interlayer distance 
in the WKB approximation for the interlayer tunneling.  This would lead
to the relation $T_{c} \propto n_{s2d}/m^{*} \times 1/c_{int} = n_{s}/m^{*}$
[56] which explains the trends in Figs. 1 and 5(c).  

As illustrated in Fig. 5(c) with the blue arrow, 
the superfluid density of the cuprates 
is substantially reduced from the $T\rightarrow 0$ value before
reaching the $T_{KT}$ line, where the thermal creation of 
vortex/antivortex pairs could destroy the superconducting state
if there is no interlayer coupling.  For a substantial 3-d 
coupling, the vortex pair creation is further suppressed, 
and $n_{s}/m^{*}(T)$ can be further reduced. 
Thus a key to understanding the correlations in Figs. 
1 and 5(c) would be to identify the process which governs the thermal
reduction of $n_{s}/m^{*}$.

\section{Excitation: magnetic resonance mode as a roton analogue}

In bulk 3-d superfluid $^{4}$He, the transition temeprature
$T_{c}$ = 2.2 K is 
determined by the phonon-roton dispersion relation
shown in Fig. 6(a) [60].  
The system loses superfluidity with this mechanism 
well below $T_{BE}$ = 3.2 K.
Although its energy $\hbar\omega_{rtn}$
corresponds to about 
4$k_{B}T_{c}$, the thermal excitation to the roton minimun
has a large weight in the phase space thanks to a substantial momentum 
transfer of $\sim 2.0 \AA^{-1}$, 
and $\hbar\omega_{rtn}$ becomes a primary determining 
factor for $T_{c}$.  This
situation can be appreciated by looking at the linear relation 
between the $\hbar\omega_{rtn}$ and $T_{c}$ in 
bulk $^{4}$He at different pressures [61], 
shown in Fig. 6(b), although only a small change is 
measurable.

It is then natural to search for any counterpart of rotons in 
the cuprates.  Here we propose the ($\pi$,$\pi$) S=1 magnetic 
resonance mode, observed by neutrons, as excitations
playing a role analogous to that of rotons.
In Fig. 6(b), we plot $T_{c}$ versus the energy of 
this mode $\hbar\omega_{res}$,
determined by neutron scattering for various cuprate systems [15-18].  
We also include the peak energy $\hbar\omega_{A1g}$
of the A1$_{g}$ mode observed in 
Raman scattering [19-22],
which closely follows the resonance mode energy.  
We see an impressive linear dependence of $T_{c}$ on 
$\hbar\omega_{res}$ and $\hbar\omega_{A1g}$
for wide variety of cuprates.  The slopes of the linear
relationships for cuprate-resonances and He-rotons are 
different only by 20 \%\ or so.

At a glance, the magnetic resonance mode is quite different 
from rotons, since 
it does not create a liberated spin-singlet charge pair nor 
charge density modulation, except through a mechanism which we
will propose in the next section.   
However, the following argument may enlighten further similarities
between these excitations.
The roton minimum represents a soft phonon mode towards solidification 
of He, which has not 
yet been achieved, at the momentum transfer close to that of 
the ``Bragg point of the solid 
He to-be''.  So, it can be viewed as an excitation mode directly 
related to the ``yet-to-be-stabilized
ground state'' competing with the superfluid ground state.        

For underdoped cuprates, likely ground states competing 
against the superconducting 
state include the antiferromagnetic (AF) state and 
the stabilized incommensurate/stripe spin state.
Since the AF state developes around the momentum transfer ($\pi$,$\pi$),
the S=1 magnetic resonance mode can be viewed as an excitation related to this 
``yet-to-be achieved and competing ground state''.  The stripe state develops
around an incommensurate wave vector.  Recently, the downward dispersion of the
resonance mode towards this incommensurate wave vector was noticed by neutron 
studies [15,62].  So, the incommensurate/stripe state could also be viewed as  
a competing ground state connected to the resonance mode.
Indeed, in La$_{2}$CuO$_{4.11}$, the 
magnetic islands with stabilized incommensurate 
spin correlations coexist with surrounding superconducting 
and non-magnetic regions [25].    
So, the superconducting and stripe states should have very 
close ground state energies and be competing with each other.

An important notion about rotons is that the 
excitation is not to create the competing ground state itself,
but is to populate ``excitations'' assciated with that competing state.
The AF state or the incommensurate spin state is the spin singlet 
S=0 state.  So, the magnetic resonance mode represents 
the ``S=1 excited state'' of the ``unachieved S=0 AF state''.
This is a way to understand the S=1 excitation as an acceptable
candidate for the roton analogue.  ``Rotons as soft phonons'' in 
$^{4}$He are {\it excitations\/} associated with the unachieved
solidified He lattice.     

We also have to realize that the thermal excitations 
of the magnetic resonance mode
could contribute towards reduction of the superfluid density $n_{s}/m^{*}$.
This feature can be found in earlier ``spin-exciton'' theories of Norman
and co-workers [63,64]
as well as ``topological transition'' theory of Onufrieva and
Pfeuty [65].  Though the relationship to the superfluid density was not explicitely
discussed in both theories, the resonance mode is described as a particle-hole
excitation from the superconducting ground state.  Thus the excitation of 
the resonance mode should reduce $n_{s}/m^{*}$.
In other words, the superconducting ground state can 
be destroyed via excitations 
associated with
the (yet-to-be-achieved) spin ordered state in the same way as those associated 
with charge ordered states.  
Now we can view the magnetic resonance mode in the cuprates as the excitation 
analogous to rotons in superfluid He as well as ``magneto-rotons'' in fractional
quantum Hall systems [66].  

In the cuprates, the resonance mode energy is related to the ``distance from 
the competing ground state''.  Thus $\hbar\omega_{res}$ increases with 
increasing doping as the system moves farther away from the AF ordered state.
It has been difficult to observe the resonance mode by 
neutron scattering in the La214 systems.
However, if we adopt the Raman A1$_{g}$ results, 
$T_{c}$ of the 214 system also follows the linear relation
with the mode energy, as shown in Fig. 6(b). 
The resonance mode appears more clearly in the 
low-energy dispersing branch
at the incommensurate position in LSCO [62].  In Fig. 5(b),
we include the energy at ($\pi-\delta,\pi$)
in YBCO [17] and LSCO [62], which exhibits an even better agreement with
the slope of rotons.  

In theoretical works for the resonance mode, we often
find statements such as ``the reason is unknown for why the mode energy scales 
with $T_{c}$''.  We claim here that $T_{c}$ is determined by the mode energy
via thermal excitations of the mode which reduce superfluid density.
The actual situation can, however, be somewhat more complicated since not only 
the resonance mode but also the ``continuum'' excitations 
(or the B1$_{g}$ Raman modes)
would contribute towards reduction of the superfluid density in the cuprates,
and the energies of the resonance and continum modes 
are very close to each other. 

\section{Reconciliation of Raman and neutron modes}

Raman scattering usually probes spinless charge excitations
at nearly zero net momentum transfer.  How, then, can we
explain the Raman A1$_{g}$ mode to detect the S=1 magnetic resonance mode
which involves spin-flip and momentum transfer of ($\pi$,$\pi$) ?
This is still a remaining open question.  Possible but seemingly unlikely 
answers could include a spin-flip Raman mode involving
spin-orbit coupling, detecting 
spin fluctuations near k = 0 within the 
``shadow AF zone'' dispersion corresponding to ($\pi$,$\pi$).
A standard two magnon Raman mode cannot explain the same energy transfer of 
the Raman and neutron peaks, since in that case the Raman $\hbar\omega_{A1g}$
should have been a factor of 2 larger than the neutron $\hbar\omega_{res}$.
In superfluid $^{4}$He, simultaneous creation of two rotons is known to give 
a strong Raman signal [67], assuring momentum conservation with nearly 
opposite momentum transfers of the two rotons.

Recently, STM studies [11,68,69] found evidence for inelastic charge
modulations spreading with the periodicity of 4 lattice 
constants, which could provide yet another/relevant competing
ground state.  Dung-Hai Lee and co-workers [70], and 
some other authors [71] proposed a picture involving a 
roton minimum for this charge density modulation.
In general, any such charge or spin excitations 
relevant to competing ground states could play
a role similar to the resonance mode as discussed here.
However, we shall note that the contribution of 
modes with small momentum transfers  
would provide a limited effect in the reduction of $T_{c}$ due to
its limited phase space factor.  In this sense the ($\pi$,$\pi$)
resonance mode is the most
attractive candidate for the primary determining
factor of $T_{c}$.

Struggling with the question about the almost identical energies of the S=1 neutron
resonance and presumably S=0 Raman mode can bring a further possible insight
into the roton-like mode in the cuprates.  
In (La,Nd,Sr)$_{2}$CuO$_{4}$, which shows stabilized stripe
spin and charge modulations, neutron scattering [72] found magnetic 
incommensurate satellite Bragg peaks (red close circle in Fig. 6(c)), 
away from the AF zone center ($\pi$,$\pi$) by a distance 
$\delta$ in momentum space due to spin modulation. 
Also observed simultaneously are the the lattice deformation peaks at a
distance 2$\delta$ away from the lattice zone center (blue closed circle
in Fig. 6(c)), due to the 
charge modulation.  
Since the cuprate systems exhibit a preference of $\delta$
being $2\pi/8$, a spin/charge ordered stripe state with 
$\delta = 2\pi/8$ can be
a natural strong candidate of the ``competing ground state''.  
Having the periodicity 2$\delta = 2\pi/4$, the charge modulation observed by
STM can well be a manifestation of this stripe modulation or similar 2-d
incommensurate correlations in a dynamic response.

Suppose the superconducting ground state wins in the competition against 
this incommensurate state.  This will lift the energy of both the magnetic and charge
satellite Bragg peaks to finite energy transfers and create the magnetic resonance mode 
near the magnetic satellite as well as the charge modulation mode near
the charge satellite.  These spin and charge modes should co-exist simultaneously.
The energy of these modes is determined by the energy difference between 
the superconducting state and the stripe state.  Therefore, we would expect
the {\bf same energies\/} for the charge and spin modes, but these 
modes exist, of course, with different momentum transfers.    
This special situation creates a novel roton-like excitation in cuprates 
with double charge and spin minima, as illustrated in Fig. 6(c).
We shall call this as a ``hybrid spin/charge roton''.   

This double-minima hybrid roton provides a natural resolution for the 
neutron vs. Raman conflict.  The 
S=1 magnetic neutron mode is due to the spin branch
while the Raman results can be 
understood as the manifestation of the charge/lattice branch,
which would show up as the S=0 response near the zone center.
It is then natural to find these modes with the same energy transfer,
as actually observed in experiments shown in Fig. 6(b).

Rotons are usually viewed as representing the density modulation
in neutral superfluids, and charge modulation in charged superconductors.
The spin mode in cuprates, however, can exist only if the dynamic combined
spin-charge modulation assures that this mode is associated with the 
``to-be-achieved'' static spin-charge modulated ground state.
Therefore, the existence of the spin mode
directly implies simultaneous dynamic charge modulation.  
This argument gives the basis for why the excitation of the magnetic
resonance mode can contribute to the reduction of the 
superfluid density $n_{s}/m^{*}$, or in other words to the destruction of
superconducting order parameter, in the cuprates. 

In charged superfluids and/or metals, the $k=0$ response of the density 
fluctuations corresponds to plasmons, which usually exist at a rather 
high energy transfer.  Then, one tends to presume a  
steep reduction of the charge-roton branch towards the 
$2\delta = 2\pi/4$ minimum from the zone center, instead of a rather modest energy reduction
shown in Fig. 6(c).  The in-plane plasma frequency $\omega_{p}$ at $k=0$ for 
the optimally doped YBCO system, determined by 
optical conductivity [73], has an 
energy scale at least one order of magnitude higher than that of the 41 meV mode energy.   
This energy scale is also represented roughly by $T_{F}$ obtained by the superfluid density
in Fig. 2, as the product of $\lambda$ and $\omega_{p}$ is 
the light velocity.    
These energies 
represent ``real responses of the superconducting ground state''.
 
In the parent Mott insulator, all the charges are localized,
and thus the plasma frequency is pushed down to zero, 
since the effective mass is infinitely large.
Just as the magnetic mode energy represents the low-energy
spin waves of the {\it yet-to-be achieved Mott insulator AF state\/}, 
the charge branch in Fig. 6(c) should be considered 
for a {\it hypothetical situation\/} with a charge injected 
into a Mott insulator,
somewhat comparable to the case of 
photo-induced conductivity experiments of a parent insulator system.
Then, one might understand the very low energy scale, which 
could be related predominantly to  
nearly localized charges.
Another way to view the roton branches: suppose
we temporarily created a small island region of the ``un-achieved competing ground state'',
corresponding to a vacancy in the ``swiss cheese'', 
by supplying the finite energy transfer to the superconducting system, 
and study {\it excitations\/} within that island.  Then we will 
find all the roton-like branches.
This argument explains why the energy of the charge mode at the zone center
$k=0$ can be identical to that of the magnetic mode at ($\pi$,$\pi$).
More formally, this spin-charge coupling would provide
terms analogous to the case of strong spin orbit coupling, 
giving some qualification to the ``seemingly unlikely''
argument of spin-flip Raman scattering described in the beginning of this section.
However, one has to wait for further accumulation of experimental studies to 
clarify details of the ``charge branch'' near the zone center 
and its interplay with the lattice and 
spin degrees of freedom.     

\section{Phase separation in the overdoped region}

In 1993, our group [24] and a European $\mu$SR team [30] independently found a 
very strange variation of $n_{s}/m^{*}$ in overdoped Tl2201.
As shown in Fig. 1, $n_{s}/m^{*}$ decreases with increasing carrier doping,
yet conductivity studies imply no anomaly in $m^{*}$.  We presented a 
picture involving phase separation into superconducting and normal 
metal ground states [24,7], while authors of ref. [30] interpreted 
this in terms of pair-breaking 
scattering.  Co-existence of superconducting and normal 
charges at $T\rightarrow 0$ in Tl2201 also shows up in specific heat
measurements by Loram {\it et al.\/} [74] shown in Fig. 7(a).  
Upon overdoping from the highest $T_{c}$ sample in the nearly
optimum region, the ``un-gapped'' response develops in the 
T-linear term $\gamma = C_{el}/T$ of the electronic
specific heat $C_{el}$.  As shown in Fig. 7(b), the ``gapped'' fraction,
presumably representing the volume fraction of the superconducting region,
closely follows the variation of the superfluid response $n_{s}/m^{*}$
obtained from $\mu$SR [24].  

Further evidence of co-existing gapped and ungapped ground states can be
found in the height $\Delta C$ of the specific heat jump at $T_{c}$.  
In BCS superconductors,
the height $\Delta C$ at $T_{c}$ is proportional
to $T_{c}$, as illustrated in Fig. 7(c).  In the $\gamma$ versus $T$ plot,
ideal BCS systems with different $T_{c}$'s should exhibit the same 
jump height $\Delta C/T_{c}$, as illustrated in Fig. 7(d).  Comparison of 
Figs. 7(a) and 7(d) demonstrates that the superconducting 
condensation in overdoped Tl2201 occurs only in a finite volume fraction.
Thus the overdoped system is fundamentally different from standard
BCS superconductors where all the normal-state charge carriers 
participate in the superfluid $n_{s}$ once the energy gap is developed
around the entire Fermi surface.  The jump height $\Delta C/T_{c}$ follows the 
trend of $n_{s}/m^{*}$ and the gapped volume fraction, as shown in Fig. 7(b),
further reinforcing this argument.

Similar behavior in the other overdoped cuprates have been found by subsequent
$\mu$SR measurements in YBCO with (Y,Ca) substitution [75] 
and in CaLaBaCuO [31], and torque
measurements of Hg1201 [76].  Other signatures of 
this anomaly include the un-gapped response in optical conductivity [77]
in optimal to overdoped YBCO, 
and the recombination time of photo-induced anti-nodal quasiparticles
which becomes independent of laser excitation power in the overdoped
region of Bi2212 [78].  All these results indicate that the overdoped cuprate
systems have a substantial number of unpaired fermion 
carriers in the ground state.  As a condensed boson
superfluid co-existing with fermionic carriers, the overdoped cuprates
strongly resemble $^{4}$He/$^{3}$He mixture films.  Indeed the $T_{c}$ versus
$n_{s}/m^{*}$ relationship in these two cases follow very similar behaviors,
as shown in Figs. 1 and 5(b).

\section{Phase diagrams and the Nernst effect}

\subsection{phase diagrams}

In recent years, we presented a generic phase
diagram for the cuprates [79,33] shown in Fig. 8(a), which is a hybrid of the 
BE-BCS conjecture and the phase separation picture in the overdoped region.
We assume that the pairing energy, represented by $T^{*}$, vanishes
{\it within\/} the superconducting dome 
at a critical concentration $p_{c}$ in a 
modestly overdoped region.
Tallon and Loram [80] obtained
$p_{c} \sim 0.19$ holes per Cu for YBCO, Bi2212 and LSCO systems,
while
Hwang {\it et al.\/} [81] obtained $p_{c} = 0.23$ for Bi2212, 
and some other estimate [82] suggeests that $p_{c}$ could be even 
larger in LSCO, as illustrated in Fig. 8(b).
These subtle differences
may be partly due to different definitions used to derive $T^{*}$ 
from results of various experimental methods. It is also possible
that $p_{c}$ substantially varies from system to system.

In our view, the phase separation 
in the overdoped region occurs to save condensation and pairing
energies at a cost of (screened) Coulomb energy necessary for 
disproportionation of charge density [79].
Without taking into account the possibility of phase separation, 
the authors of refs. [80] and [81] argued that the 
existence of superconductivity at 
the hole concentration $x > p_{c}$
could rule out a positive role played by the pseudogap state for superconductivity.  
Our phase separation picture, however,  
presents a counter-example to this argument and shows that  
the superconducting state can exist 
at $x > p_{c}$ with $T^{*}$ representing 
the onset of an interaction {\bf necessary\/} for 
superconductivity.   

The reduction of $T^{*}$ with increasing doping is well established by 
many experimental results.  However, this is not a feature readily expected
in a general argument of BE-BCS crossover.  In fact, if the pair formation
energy did not show a strong dependence on doping, we could expect 
$T_{c}$ to rise smoothly up to the $T^{*}$ energy scale with 
increasing doping,  after which 
the system slowly crosses over to a standard BCS behavior with all the 
normal state carriers condensing in the superconducting state without
spatial phase separation.
$T_{c}$ would still come down gradually in the high-density limit due to the 
effect of retardation.
Compared to this ``standard BE-BCS crossover'', the situation in the
cuprates is different, because the ``origin of pairing interactions''
seems to die away as the doping progresses.

\subsection{Nernst effect}

The results of the high-frequency
superconducting response [12] and the Nernst effect [13,14]
demonstrate the existence of vortex fluctuations 
(or ``superconducting phase fluctuations'' 
or ``dynamic superconductivity'') in a wide region of 
the normal state above $T_{c}$, as shown in Fig. 8(b) [83].  In 
thin films of $^{4}$He, the corresponding region exists above 
the superfluid temperature $T_{c}=T_{KT}$ 
up to the ``mean-field'' superfluid energy scale whose 
upper limit can be given by the lambda transition temperature 
2.2 K of the bulk $^{4}$He.  This comparison leads us
to realize that the ``Nernst region'' could have
been superconducting if the system were 
protected against various origins for the destruction of
the superconducting states, including
low dimensional aspects,
excitation of roton-like modes, etc., but
all the other necessary characters for superconductivity were
conserved.  Then, the onset temperature $T_{on}$ of the 
Nernst effect corresponds to $T_{c}$ of such a ``hypothetical 3-d
and roton-less
counterpart'' of highly 2-d cuprate systems.

In Fig. 2, we include a point (green star) 
having $T_{F}$ of La$_{1.9}$Sr$_{0.1}$CuO$_{4}$ in the horizontal
and the Nernst onset
$T_{on}$ of the same system [13,14] on the vertical axes, respectively.  
Interestingly, this point for 
``hypothetical underdoped LSCO'' lies very close to the 
$T_{BE}$ line, similar to those of bulk $^{4}$He and ultracold
$^{40}$K.  This suggests that the effect of pair overlapping in the
very underdoped region may be modest, and/or the
coherence length could be comparable to the intercarrier distance.
Indeed, the new estimate for $\xi$ from the Nernst study [14]
supports the latter picture.  

For further doping, $T_{on}$ is suppressed
because $T^{*}$ is reduced.  It is important, however, to realize
that the pair formation energy scale $T^{*}$ and the on-set of 
dynamic superconductivity $T_{on}$ represent two distinctly different
energy scales on the BE side.  Pair formation can occur
at any high temprature if a strong attractive force is provided, 
while the fluctuating superconductivity is possible only 
below $T_{BE}$.   Observation of the Nernst effect
above $T_{c}$ does not distinguish between BE and BCS type condensations,
since a thin film of a typical BCS superconductor can also exhibit 
the Nernst effect
above $T_{c}$.  The more important feature for underdoped cuprates
is the large region in the $T$-$x$ phase diagram
above $T_{on}$ but below $T^{*}$, as shown in Figs. 8(a) and (b),
where one would expect the emergence of normal but paired (2e) charge carriers
coexisting with unpaired (e) carriers.

The observation of Nernst effect above $T_{c}$ does not necessarily 
imply KT like phase-fluctuations as the primary determining factor for $T_{c}$.
In low-dimensional magnetic systems, we expect 
correlated spin fluctuations in a large temperature region 
above the 3-d ordering temperature, for both cases with or without a KT
transition.  Analogous to this, 
one can expect the Nernst effect for any superconducting system 
with strong anisotropy, where
$T_{c}$ is reduced from the 3-d values 
but the transition temperature is still governed by the 
anisotropic coupling strength.  Furthermore, excitations of roton-like
modes would create liberated vortices and thus also contribute towards
the Nernst signal.  

\subsection{determining factor for $T_{c}$}

In Fig. 1, the trajectory for the La214 systems
exhibits ``early departure'' from the linear line of YBCO
and other cuprates.  La214 systems have 
a particular closeness to competing magnetic ground
states as demonstrated by the instability against spin/charge
stripe formation near the 1/8 concentration.
Then, we can ascribe the reduced $T_{c}$ (about a factor 2 
smaller than that of YBCO for a given
$n_{s}/m^{*}$) to the reduced 
(and likely broadened) resonance mode
energy. 

In bulk superfluid $^{4}$He, $T_{c}$ is related to the 
particle density, mass, and the roton energy.
The roton energy provides a proper account
for the reduction of $T_{c}$ from $T_{BE}$ 
due to the finite-boson-size and interaction effects.
In thin-films of He, $T_{c}$ is determined primarily
by the particle density and mass via the KT transition,
since the roton energy there would be much higher than $T_{KT}$.
In underdoped cuprates, as we have seen, 
$T_{c}$ seems to be related to the 
doped hole density and mass as well as to the resonance mode
energy.  The mode energy will become higher with hole doping
as the system becomes distant from the AF state, while
the increasing superfluid density would also help lowering
the energy of superconducting state. Both of these would 
help increasing $T_{c}$ approximately linearly with 
the doping $x$.  

However, the decrease of $T^{*}$ with increasing $x$ implies 
that the origin of superconducting pairing is destroyed
gradually with increasing hole density $x$. In other words, the distance
from the competing AF Mott insulator state helps 
increase the roton-like resonance energy while 
decreasing the coupling strength of superconducting pairs.
This situation can be expected if the AF interaction provides
the very origin of pairing.

\section{A microscopic model for pairing mediated by spin fluctuations}

\subsection{motion of a charge resonating with AF spin fluctuations}

Stimulated by this possibility, we now develop a microscopic model 
for a unique motion of a charge in nearly
AF but actually metallic cuprate systems.  Let us consider how a charge
(hole) motion can avoid creating frustration in the region with a short-range
and dynamic AF correlations.  We start with a story for a single hole
illustrated in Fig. 9(a).  The surrounding AF configuration is fluctuating
with the frequency $\omega_{AF}$, which implies that all the surronding
Cu spins change their directions within half a period $\pi/\omega_{AF}$.
If the charge hops to the adjacent site at this half period, then 
there is no extra spin frustration in the surroundings, see Fig. 9(b).
During the next half period of spin fluctuations, the charge could 
further hop to the next adjacent site, as Fig. 9(c).  
If the charge motion occurs this way,
sequenced with the spin fluctuation, extra spin frustration can be avoided. 
Charge motion of any other (higher) 
frequency would make a mis-match with the spin lattice,
and thus cost extra energy for creating frustrated magnetic bonds.

For a charge proceeding towards the Cu-O-Cu bond direction [i.e., the ($\pi$,0)
direction in the reciprocal space], this implies that within the full period
of spin fluctuations, the charge proceeds by 2 lattice constants.  
This charge motion has the wavevector
$2\pi/2a$, corresponding to that of the ``antinodal charge'' at the 
($\pi$,0) point of the Brillouin zone.  So, the antinodal charge
can proceed comfortably in the nearly AF environment, if its 
frequency (kinetic energy)
``resonates'' with the frequency of the AF spin fluctuations.
A specially strong spin-charge coupling can be expected for the case with
$\hbar\omega_{antinode}=\hbar\omega_{AF}$.

This situation can be viewed as a charge motion creating the AF fluctuation with 
the frequency $\hbar\omega_{AF} = k_{B}T_{F}$, just as a motion of a charge
in BCS superconductors creates a phonon.  Analogous to BCS coupling mediated by a phonon, 
the AF fluctuating environment created by the first charge motion
in the cuprates can be appreciated by the arrival of 
the second charge.  This would support coupled 
charges with a strong attractive interaction 
mediated by spin fluctuations.  

In reciprocal space, this implies connecting the antinodal charges with AF fluctuation
as illustrated in Fig. 10(a).   
An initial charge at ($0,-\pi$) creates a ($\pi,\pi$) AF fluctuation and
ends up in ($\pi,0$) charge in the final state.  The same AF fluctuation
can be absorbed by another charge having initial state of ($0,\pi$)
which is scattered into ($-\pi,0$).
The large energy gap in the antinodal positions, opening below $T^{*}$, would
be a manifestation of the strong coupling of antinodal charges via this
mechanism.
In this case the spin fluctuations are analogous to phonons in BCS coupling
as a ``virtual'' boson in the ``scattering'' process.
However, the direct coupling 
between the ($\pi,0$) and ($-\pi,0$) charges is not
supported since the AF ($\pi,\pi$) wavevector does not connect antinodal
charges with opposite momentum directions. 

\subsection{coupled nodal charges and the resonance mode as a pi-meson analogue}

Let us now consider the motion of a single nodal charge at ($\pi/2,\pi/2$) in the 
Brillouin zone.  To achieve propagation along the nodal direction, this 
charge should hop, after the first hop in Fig. 9(b), towards the perpendicular
direction from the first hop, leading to the state shown in Fig. 9(d) after
one period of the AF spin fluctuation.  In this case, the charge has moved 
the distance corresponding to the diagonal of the unit CuO$_{2}$ square lattice,
inferring the corresponding wavevector ($\pi,\pi$), twice as large
as that of the nodal carriers.  Due to this complete mismatch, the single
nodal carrier cannot appreciate benefit of the ``resonating motion''.
However, the ($\pi,\pi$) AF fluctuation can connect two 
nodal charges, with opposite momentum
directions, as shown in Fig. 10(b).  
This process provides a stable
``bound state'' of two nodal charges, with the AF spin fluctuation 
playing a role of the Yukawa meson in the binding of hadrons in nuclei.
  
The energy of the AF fluctuation mediating this direct coupling
is related to the size of the boson, similarly to the 
pion mass related to the size of deuteron nucleus, and also
similarly to the energy gap related to the coherence length in
BCS superconductors.  We note 
that the resonance mode energy of $\sim 41$ meV is just
appropriate for creating a boson having the size of 
coherence length $\sim 10-20$ \AA\ in the cuprates.
Thus we suggest the possibility that the magnetic resonance mode
can be the pair mediating boson for nodal charges.

\subsection{nodal vs. antinodal carriers and ARPES coherence peak}

As discussed in earlier sections, we also expect the existence of 
``the charge branch'' of the hybrid spin/charge roton, 
and/or higher energy charge fluctuations, which could 
cause scattering of nodal charges with a rather small momentum change
within the ``nodal hole pocket'', as shown in Fig. 10(d), 
to contribute to an additional scattering process for 
an attaractive interaction.
Nearly AF spin fluctuations can also contribute to this type
of scattering process between the two different nodal hole
pockets across the zone center, in addition to the direct binding.

When the charge branch excitation couples two antinodal particles
near the ($0,-\pi$) points, one of them can be viewed as a ($0,+\pi$)
particle in an Umklapp process,
as illustrated in Fig. 10(c).   Then the charge branch
roton can be the direct ``binding boson'' of antinodal charges, similarly
to the spin branch for nodal charges.
Since the ARPES measurements probe spin-non-flip S=0 charge processes,
and since the coherence peak selectively appears in antinodal responses,
the ARPES coherence peak can be viewed as the process of detecting/liberating
the charge-branch roton via energy supplied by a photon, analogous to the Raman A1$_{g}$ mode.  
Figures 10(e) and (f) compare this
process with neutrons creating the magnetic resonance mode for
nodal particles.  In these ``liberation processes'' of binding
bosons, we expect the intensity to be proportional to the number
of condensed bosons.  This is similar to pion creation 
with accelerated particle beam, in which the intensity is proportional to the 
number of nucleons existing in the production target.  This argument explains why the 
ARPES coherence peak and neutron resonance peak both appear
with intensities proportional to the superfluid density $n_{s}/m^{*}$.
 
The interplay between the ``scattering'' process (Fig. 10 (a) and (d)) and the 
``binding'' process (Fig. 10 (b) and (c)) is reminiscent of 
the Feshbach resonance [40,84] appearing in the BE-BCS crossover of 
ultracold atoms.
Note that some of these
processes in Fig. 10, such as the binding ones in (b) and (c), 
may take place simultaneously / cooperatively as 
combined spin/charge fluctuations, as discussed earlier.
In the Hamiltonian of such coupled processes, the spin fluctuation 
operator (green line in Fig. 10)
and the charge fluctuation operator (light blue line) might appear
together, operating simultaneously on the wave function composed of a 
combination of the nodal and antinodal wavefunctions.
This strong spin-charge coupling may not be confined
to the case of the low-energy hybrid roton in (b) and (c), but could
also exist in scattering processes with higher
energy fluctuations shown in (a) and (d).  
Then we expect high energy charge fluctuations to appear with the 
same energy transfers as that of  
the high energy AF spin fluctuations spreading over the energy region up to
about 100-200 meV or so.  The inelastic charge modulation recently 
observed by STM appears exactly in this energy region with the momentum
transfer close to that of the charge branch of the hybrid roton.
Thus, this STM signature can be viewed as a manifestation of 
the counterpart of the AF spin fluctuations.  Other possible 
signatures of the corresponding dynamic charge modulations include
the line profile 
of ARPES antinodal intensities near 100 - 150 meV, and
the ``mid-infrared reflection'' observed in optical conductivity
measurements in this energy region.

Being originally an antiferromagnetic
localized spin, the antinodal carriers may have a rather heavy
effective mass, and thus may have a limited contribution in 
the superfluid spectral weight $n_{s}/m^{*}$.  Nodal charges
are directly originating from doped holes, and known to support  
a rather high normal state conductivity, suggestive of 
a light mass.  It looks as if the antinodal
charges are helping the propagation of nodal pairs
by creating large energy gaps, while the main superfluid spectral weight
comes from the nodal pairs.  This picture gives a natural
explanation for $n_{s}/m^{*}$ being nearly proportional
to the doped (mobile) carrier concentration.  
Our model of coupled spin/charge operators suggests that the 
nodal and antinodal responses are strongly coupled, and cannot
exist separately.  Detailed roles and interplays
of these two different regions of the Fermi surface are, however, 
yet to be clarified by future research.    

\subsection{retardation and phase diagram}

In Fig. 3(b), we proposed the BE-BCS crossover region to have
$k_{B}T_{F}$ comparable to the energy of the pair mediating boson
$\hbar\omega_{B}$ in a rather arbitrary 
discussion through the ``retardation'' concept.
If we assume that the ``optimal doping'' region of the 
cuprates represents the crossover region, this argument
suggests that $k_{B}T_{F} \sim 2,000$ K of the optimally
doped YBCO could represent the energy scale of the pair mediating
bosons.  This crude estimate is consistent with spin
fluctuations as a pair mediator [56], since $\hbar\omega_{AF}$ 
develops over the energy region up to the AF exchange interaction 
$J \sim 1,200 - 1,500$ K.  

Furthermore, our argument of ``resonant charge motion''
can work only when the charge energy scale does not exceed
the spin fluctuation energy scale.  
Doping further carriers in the overdoped region would
increase the energy scale of doped normal holes exceeding the 
spin fluctuation energy scales.  Then we cannot
expect any more ``resonant charge motion'' nor ``pairing
mediated by spin fluctuations''.
This provides a reasoning from the microscopic model for
why the crossover region appears at $k_{B}T_{F} \sim \hbar\omega_{AF}$,
and why superconductivity in the overdoped region becomes
progressively weaker with anomalous coexistence of paired and 
unpaired charges.
 
\section{Discussions and conclusions}

In our model, the closeness to the competing
ground state can be a factor to reduce 
$T_{c}$ (by reducing the resonance roton-minimum
energy) and thus destructive to superconductivity, 
as well as a factor necessary to superconductivity
in supplying mediating bosons in scattering and binding
processes. 
This dual role of the competing ground state
provides the very origin
of the ``dome like phase diagram'', and 
also determines the distance of the maximum $T_{c}$ from 
the $T_{BE}$ line in Fig. 2.  Although various
different non-cuprate superconductors in Fig. 2 could have
different ``competing ground states'', a similar 
situation might exist in most of the systems
originating from Mott insulators,  
which could in the future provide some account of why their $T_{c}$
is limited to an apparently universal maximum
value of $T_{c}/T_{F} \sim 0.05$ in 
Fig. 2.  

There have been two different schools of thought in 
explaining the superconductivity of cuprates.  One assumes
the condensation of pre-formed pairs in the underdoped 
region, which is somehow perturbed in the overdoped region.
The other starts from ``strong'' superconductivity in the 
overdoped region and views it to be gradually destroyed
with decreasing carrier density in the competition against
other (presumably more magnetic) ground states in the underdoping
region.  Although still based on the 
former view point, the present picture qualitatively describes
how to account for the effect of ``competing ground states'',
and reconciles some conflict between these two views.

In theoretical discussions of HTSC or 1-dimensional systems,
``spin-charge separation'' is often the key underlying
concept.  Our model of charge propagation, in resonant with
spin fluctuations, provides a new way out of the spin-charge
separation problem and out of associated spin frustration.  
We pointed out several observations as evidence for 
extremely strong spin-charge coupling, which leads
to the superconductivity of the cuprates.
The strong involvement of the charge degrees of freedom
in pairing could help interpretation of the
isotope effect [85] and other signatures of  
lattice involvements, such as the ``kink'' in the electronic
dispersion found in ARPES studies [86].

In summary, we have elucidated the role 
of superfluid density via accumulated $\mu$SR
results and proposed the roton-like resonance
mode as a primary determining factor of $T_{c}$.
On the underdoped side, these two parameters,
$n_{s}/m^{*}$ and $\hbar\omega_{res}$ 
conspire to determine $T_{c}$, in a way
very much similar to that in superfulid bulk $^{4}$He.
In the 214 system, close to competing ground states,
the strong effect of reduced $\hbar\omega_{res}$
causes the early departure of the points in Fig. 1
from the nearly linear trend of other systems.
We proposed a few new pictures including: 1) ``hybrid spin-charge
roton'' to reconcile selection rules of
neutron and Raman modes;
2) a microscopic model which explains how
the charge motion and AF spin fluctuations can be
coupled, in a sequential resonanting way, 
to create an attractive interaction among
antinodal and nodal charges;
3) a possibility that the magnetic resonance mode plays
a role of bonding bosons, analoguous to pi-mesons, which
bind two nodal charges; and 4) the ARPES coherence peak
and Raman A1$_{g}$ mode can be viewed as manifestations of the 
charge branch of the hybrid spin/charge roton. 
We also demonstrated the robustness of 
the superconductivity in the cuprates against
spatial heterogeneity, analoguous to the 
superfluid He films, as an essential 
feature which cannot be expected in BCS superconductors.
Hopefully, these pictures elucidate the fundamental importance of 
the BE condensation concept for the cuprates and provide useful conceptual
frameworks for understanding their unique condensation and pairing aspects.

\section{ACKNOWLEDGMENT}

The work at Columbia has been supported by the NSF 
DMR-0102752 INT-0314058 and CHE-0117752 (Nanoscale Science and Engineering
Initiative).  Some of the ideas in this paper were developed
during the author's stay in IMR, Tohoku Univ., Japan, as
a Visiting Professor during January - March, 2003.
The author wishes to thank Profs. Bob Laughlin and 
Naoto Nagaosa for useful discussions.
The author has had an intuitive 
view to interprete the magnetic resonance
mode as a ``roton-like'' excitation for the past three years
or so, based on the energy analogy shown in Fig. 6(b).
When he confessed this idea as a wild speculation
to these two scientists in the spring of 2003,
they kindly pointed out several difficulties of this speculation.
Struggles to resolve some of them led the author
to develop the picture detailed in this paper.
In various stages of the development of thoughts
in this paper, discussions with 
Prof. Oleg Tchernyshyov were very illuminating.
The preprint regarding the STM work of
the 4 lattice-constant charge modulation, given by Prof. J.C. Seamus Davis,
was also very useful.

\vskip 1.0 truecm
\onecolumngrid
\newpage
\vfill \eject
\newpage
\begin{figure}[h]

\begin{center}
\vskip 1.0 truecm
\includegraphics[angle=0,width=5.5in]{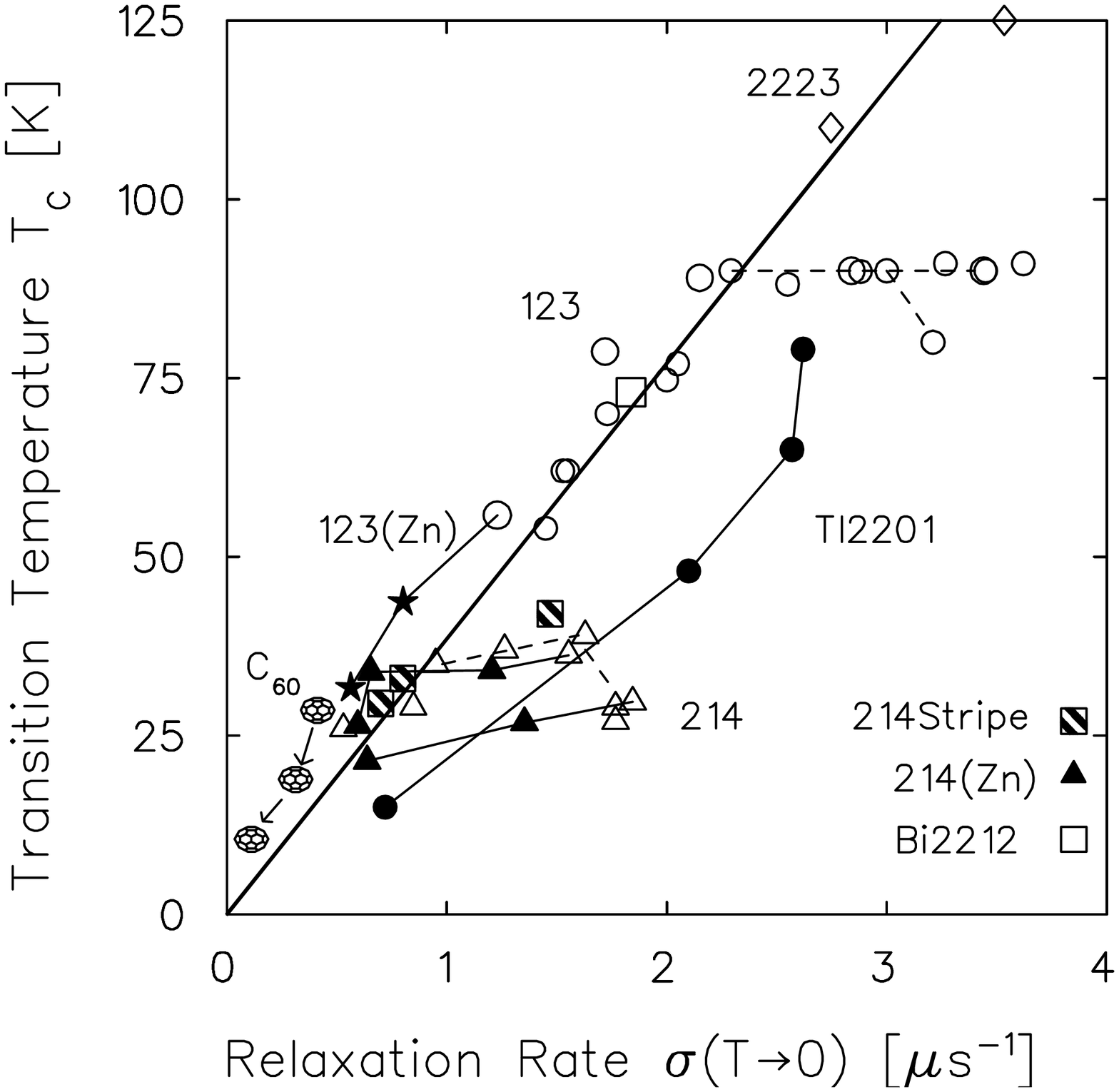}
\vskip 1.0 truecm
\label{Figure 1.} 

\caption{\label{Figure 1.}
Muon spin relaxation rate $\sigma\propto n_{s}/m^{*}$ at $T\rightarrow 0$ 
from various high-$T_{c}$ cuprate superconductors [4,5,8,24-27] and 
A$_{3}$C$_{60}$ systems [35,36] plotted against the superconducting
transition temperature $T_{c}$.  The points for HTSC with open symbols
represent simple hole doped systems, while closed triangles
are for (Cu,Zn) substitution [8], ``stripe'' symbols for
systems with formation of island regions with incommensurate
static spin modulations [25], and closed circles for overdoped Tl2201.}
\end{center}
\end{figure}
\vfill \eject
\newpage

\begin{figure}[h]

\begin{center}

\includegraphics[angle=90,width=5.5in]{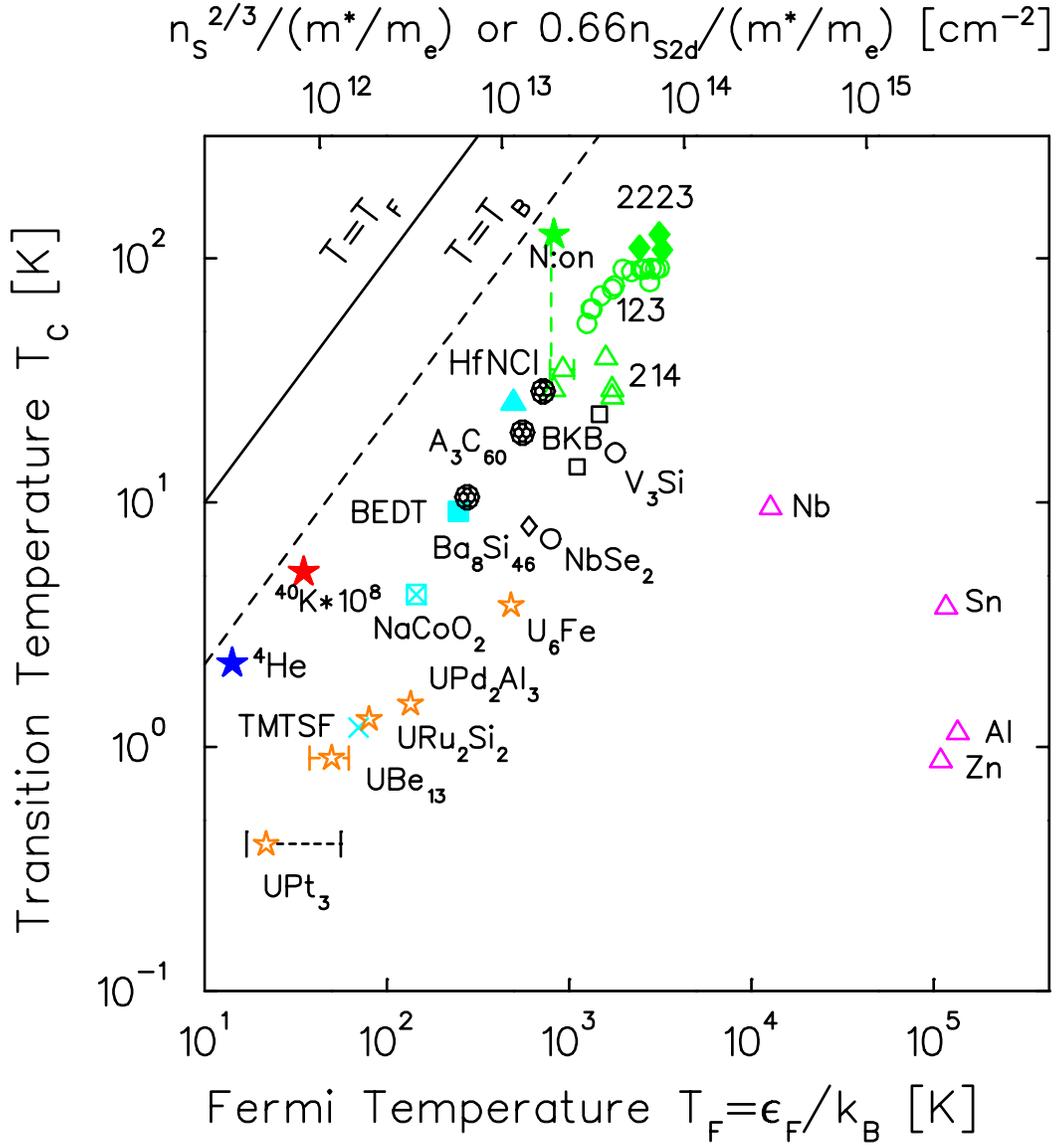}
\vskip 1.0 truecm 

\label{Figure 2.} 

\caption{\label{Figure 2.} 
Plot of $T_{c}$ versus the effective Fermi temperature
$T_{F}$ obtained from the superfluid response $n_{s}/m^{*}$
of various superconducting systems, 
first attempted in ref. [5] in 1991, and updated including
results from refs [4,5,35-39].  
We see an empirical upperlimit $T_{c}/T_{F} \sim 0.05$ 
for superconducting systems.  Also included are the corresponding
points for the superfluid $^{4}$He (blue star) and the ultracold $^{40}$K [40]
in the BE-BCS crossover region (red star; with $T_{c}$ and $T_{F}$
both multiplied by 10$^{8}$).  The $T_{B}$ line shows the 
BE condensation temperature for the 
ideal bose gas of boson density $n_{s}/2$ and 
mass $2m^{*}$.  The green star represents the onset 
temperature $T_{on}$ of the 
Nernst effect, shown in Fig. 8(b), for La$_{1.9}$Sr$_{0.1}$CuO$_{4}$ [13,14],
which represents the case 
for ``hypothetical 3-d and roton-less underdoped LSCO''.}
\end{center}
\end{figure}

\newpage

\begin{figure}[h]

\begin{center}

\includegraphics[angle=0,width=5.5in]{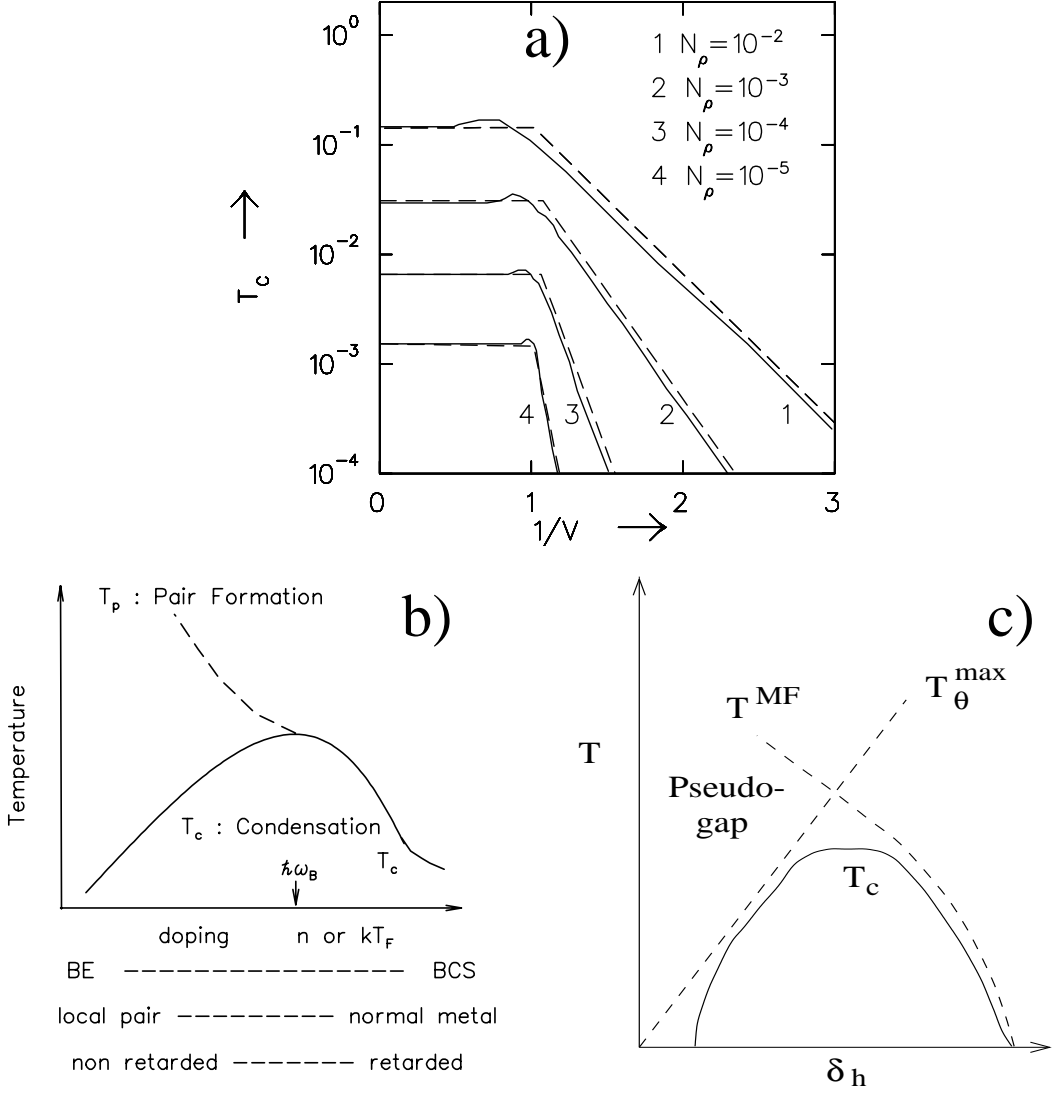}
\vskip 1.0 truecm

\label{Figure 3.} 

\caption{\label{Figure 3.}
(a) Transition temperature $T_{c}$ plotted against 
inverse of the normalized attractive 
coupling strength $V$ in the BE-BCS crossover
region obtained by Nozi\`eres and 
Schmitt-Rink [41] in 1984 for various particle
densities.  Since $V$ is defined with parameters including $N_{p}$, 
the $N_{p}$ dependence for fixed interaction strength cannot be
directly obtained from this figure.
(b) The BE-BCS crossover picture proposed by Uemura [6,7] in 1994 with the 
crossover region characterized by the matching of kinetic energy
$k_{B}T_{F}$ of the condensing carriers with the mediating boson
energy $\hbar\omega_{B}$.  When one identifies the pair formation 
temperature $T_{p}$
as the pseudo-gap temperature $T^{*}$, this phase diagram can be
mapped to the case of HTSC.
(c) A phase diagram based on the superconducting phase fluctuations in the
pseudogap region proposed by Emery and Kivelson [45] in 1995 as an
explanation to the relationship shown in Fig. 1.  $T^{MF}$ represents the 
mean-field attractive interaction, while the $T_{\Theta}^{max}$ 
shows the maximum temperature up to which the superconductivity 
in 2-d systems can survive against thermal excitations of phase fluctuations.}
\end{center}
\end{figure}
\newpage
\begin{figure}[h]

\begin{center}

\includegraphics[angle=0,width=5.5in]{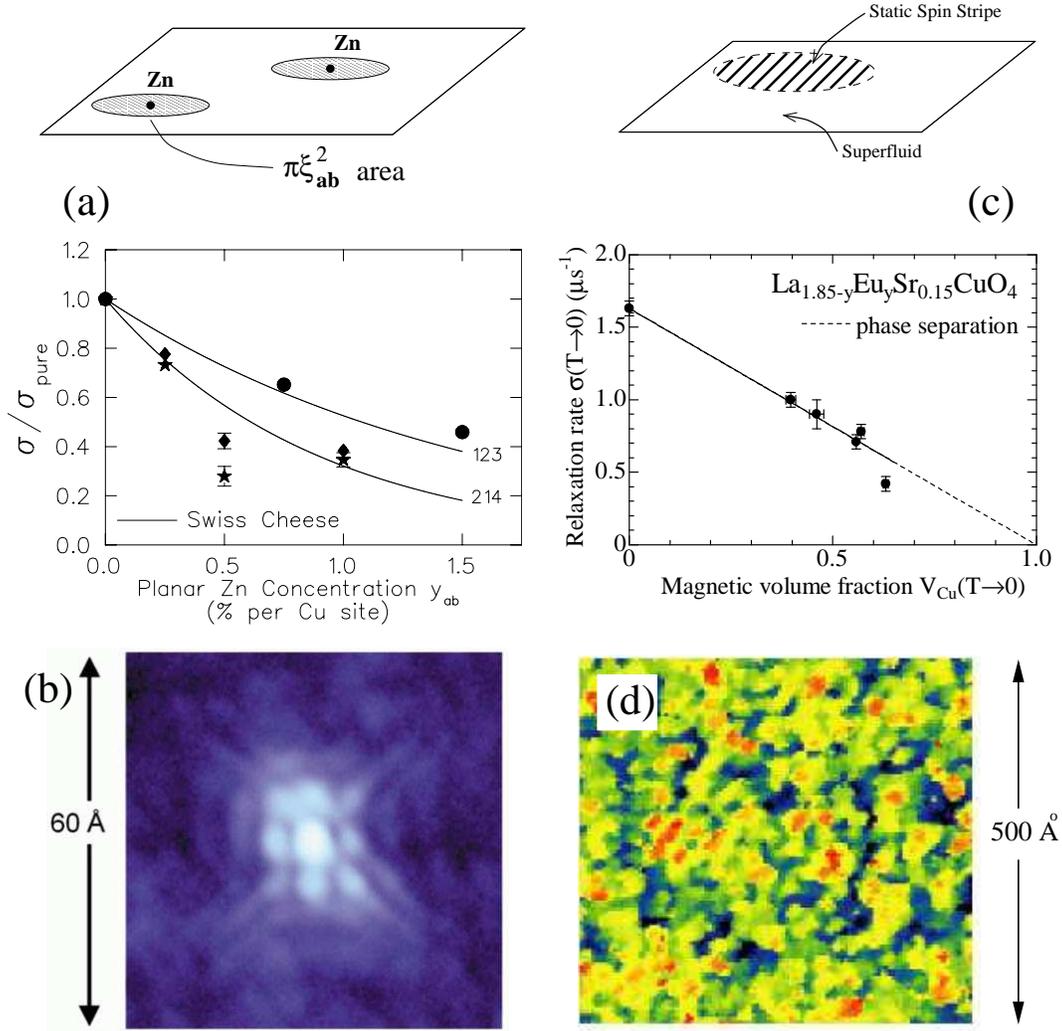}
\vskip 1.0 truecm

\label{Figure 4.} 

\caption{\label{Figure 4.}
(a) Muon spin relaxation rate 
$\sigma(T\rightarrow 0) \propto n_{s}/m^{*}$ plotted 
against the planar Zn concentration in the (Cu,Zn) substituted
YBa$_{2}$Cu$_{3}$O$_{6.63}$ (123) and La$_{2-x}$Sr$_{x}$(Cu,Zn)O$_{4}$
with $x=0.15$ and 0.20 [8].  The solid line represents 
the variation expected
for the ``swiss cheese model'' illustrated on the top, where superconductivity 
around each Zn is assumed to be destroyed in the area of $\pi\xi_{ab}^{2}$.
(b) The gapped response (blue color) versus normal state density of states
(white color) detected around one Zn impurity in (Cu,Zn) substituted
Bi2212 by STM [9].
(c) Muon spin relaxation rate $\sigma(T\rightarrow 0) \propto n_{s}/m^{*}$
in Eu substituted (La,Eu)$_{1.85}$Sr$_{0.15}$CuO$_{4}$ plotted against
the volume fraction $V_{Cu}$
of Cu moments having static magnetic order [26].  
The results demonstrate the trade off between superconducting and magnetic
volume fractions, expected in the ``magnetic island formation'' illustrated
at the top.
(d) The ``gap-map'' obtained by STM in moderately underdoped Bi2212 [10,11].
The red-yellow regions represent the area where the sharp superconducting 
gap is observed, while blue and black region has a response similar to 
those found in the ``pseudogap'' region [68] or the nonsuperconducting 
very underdoped region [69].  The characteristic length scale for 
this spontaneous formation of spatial heterogeneity is comparable to the
coherence length $\xi_{ab}$.} 
\end{center}
\end{figure}
\newpage
\begin{figure}[h]

\begin{center}

\includegraphics[angle=0,width=6.0in]{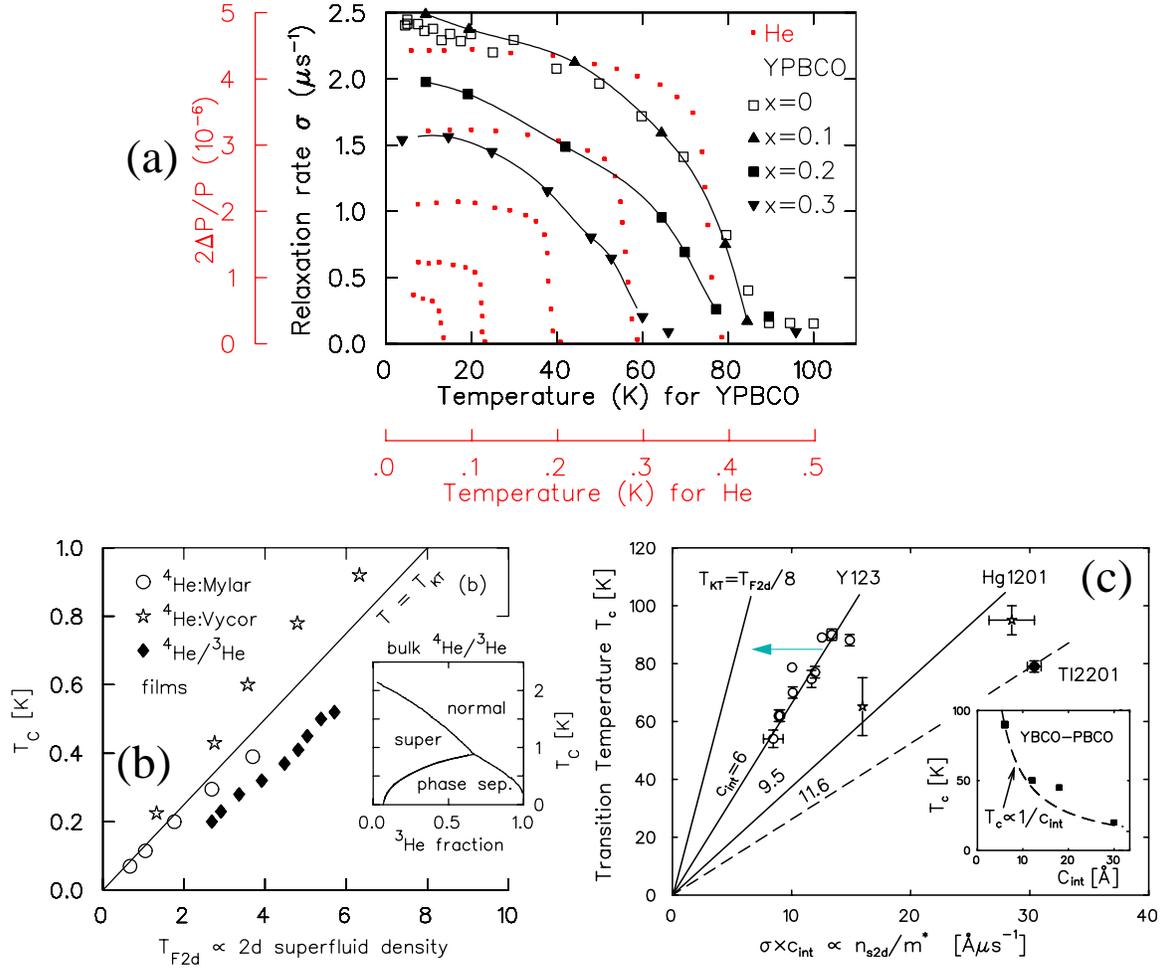}
\vskip 1.0 truecm

\label{Figure 5.} 

\caption{\label{Figure 5.}
(a) Temperature dependences of the superfluid responses
in the thin-film of $^{4}$He adsorbed on Mylar films (red points
and axes) [51] compared with those in (Y$_{1-x}$Pr$_{x}$)Ba$_{2}$Cu$_{3}$O$_{7}$
observed by $\mu$SR in unoriented [27] and oriented (multiplied with 1/1.4) [52] 
ceramic specimens. The solid lines are guides to the eye.
(b) The superfluid transition temperature $T_{c}$ versus the 2-d superfluid
density at $T\rightarrow 0$ for $^{4}$He films on regular (Mylar) [51] and
porous (Vycor) [54] media, and $^{4}$He/$^{3}$He mixture films adsorbed on 
fine alumina powders [55] where the bulk phase separation in the inset figure
is changed to microscopic heterogeneity via porous/powder media.
The linear line represents values 
of superfluid density jump expected at $T=T_{c}$
from the Kosterlitz-Thouless theory [46].  
(c) $T_{c}$ plotted against the 2-d superfluid density 
$n_{s2d}/m^{*} \propto \sigma(T\rightarrow 0)\times c_{int}$ for 
various cuprate systems
having different average interlayer spacing $c_{int}$ [56].  
The inset figure shows the $T_{c}$ variation against $c_{int}$ in 
MBE YBa$_{2}$Cu$_{3}$O$_{7}$ films where non-superconducting PrBa$_{2}$Cu$_{3}$O$_{7}$
was sandwiched with every unit cell along the c-axis direction [58].  The blue 
arrow illustrates the reduction of the superfluid density from the $T\rightarrow 0$
value towards the value expected at $T_{KT}$.  The YBCO data in (a) shows even further
reduction from this ``$T_{KT}$ jump'' [57] value as the $T_{c}$ is approached.}

\end{center}
\end{figure}
\newpage
\begin{figure}[h]

\begin{center}

\includegraphics[angle=0,width=6.0in]{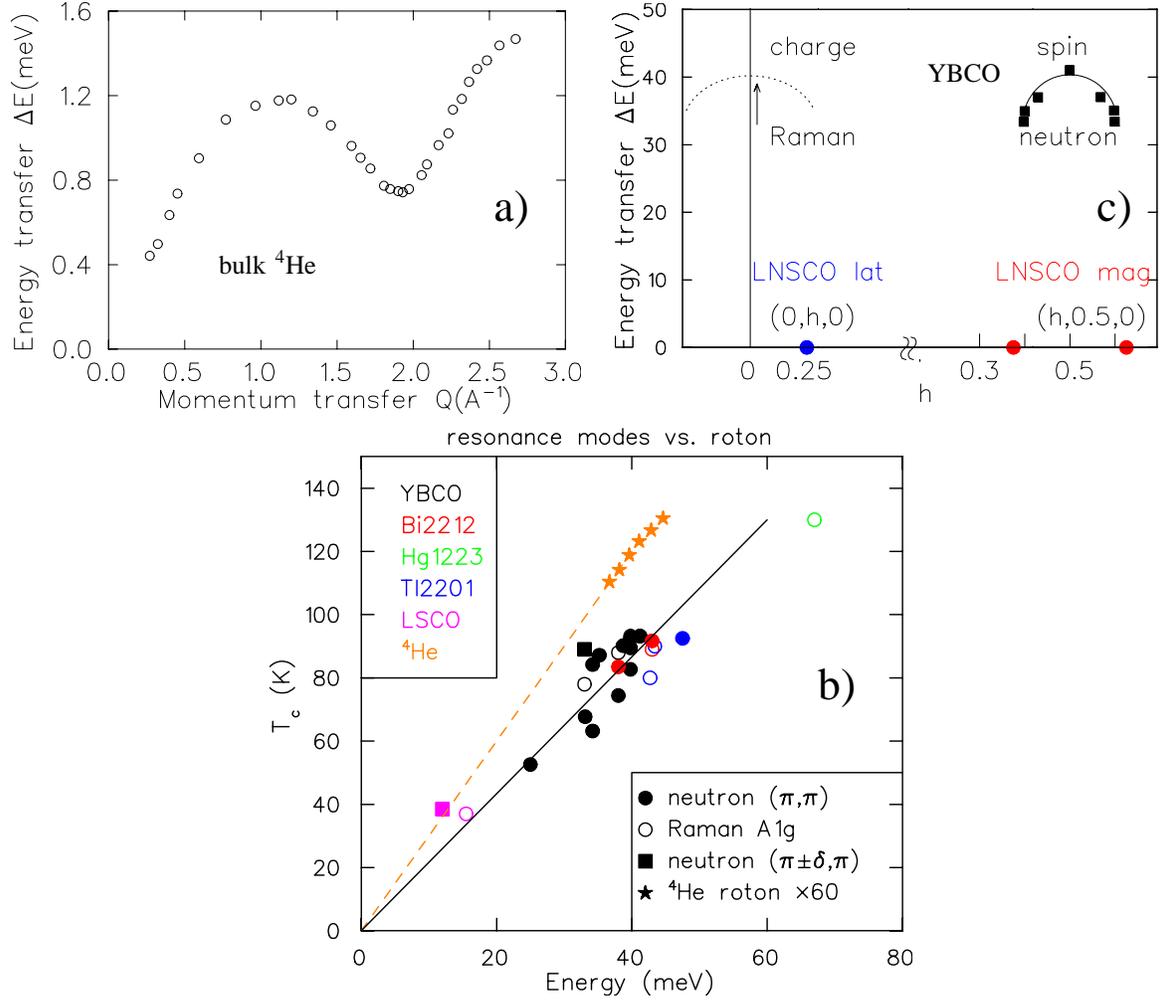}
\vskip 1.0 truecm

\label{Figure 6.} 

\caption{\label{Figure 6.}
(a) The dispersion relation of phonon-roton excitations in superfluid
$^{4}$He observed by neutron scattering [60].
(b) The plot of $T_{c}$ versus energy of the roton minimum of 
bulk superfluid $^{4}$He (values for both axes multiplied by 
a factor 60) measured under applied pressures [61],
compared with the relationship seen in 
HTSC systems for the energies of the neutron resonance mode at 
($\pi,\pi$) [15-18] and the Raman A1$_{g}$ mode [19-22].
Also included are the neutron energy transfers at the 
($\pi\pm\delta,\pi$) point in YBCO [17] and LSCO [62].
(c) The dispersion relation around the ($\pi,\pi$) resonance mode
observed in YBCO by neutrons [17] (closed circles), 
the location of the satellite Bragg peaks (red =  magnetic; blue = lattice, estimated
from the adjacent Brillouin zone)
found in the static spin/charge stripe system 
(La,Nd,Sr)$_{2}$CuO$_{4}$ (LNSCO) [72],
and the proposed charge branch of the hybrid spin/charge roton
(dotted line), to which the Raman A1$_{g}$ response in (b) is ascribed.}

\end{center}
\end{figure}
\newpage
\begin{figure}[h]

\begin{center}

\includegraphics[angle=0,width=6.0in]{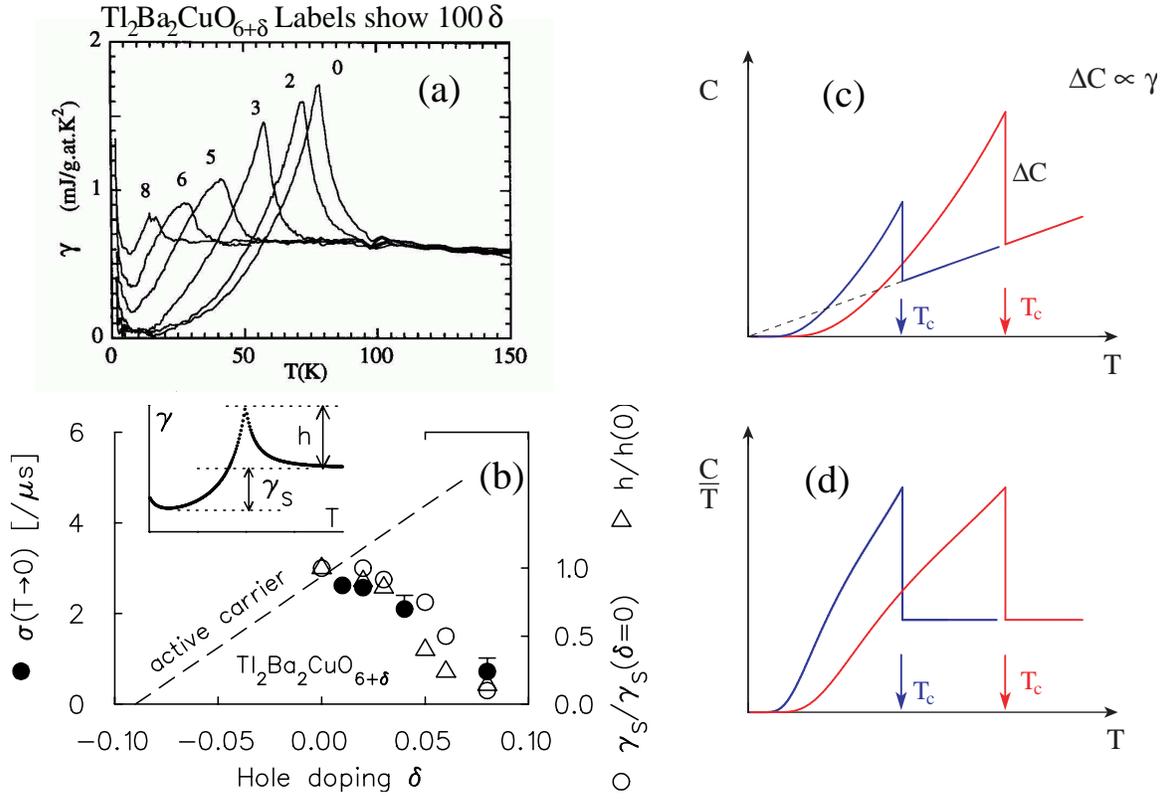}
\vskip 1.0 truecm

\label{Figure 7.} 

\caption{\label{Figure 7.}
(a) The electronic specific heat observed by Loram {\it et al.\/}
[74] in the overdoped Tl2201 systems.
(b) The variation of the $\mu$SR relaxation rate $\sigma(T\rightarrow 0)$
in Tl2201 [24] compared with the ``gapped'' response $\gamma_{S}$ (open circles) and the peak
height h (open triangles) from the data in (a).
(c) The specific heat jump in standard BCS superconductors with 
different $T_{c}$'s. (d) Same as (c) but plotted as $\gamma = C/T$, 
to be compared to (a).  
Figures (a)-(d) together provide strong evidence for 
phase separation in the overdoped cuprates developing with 
increasing overdoping.}
\end{center}
\end{figure}
\newpage
\begin{figure}[h]

\begin{center}

\includegraphics[angle=0,width=6.0in]{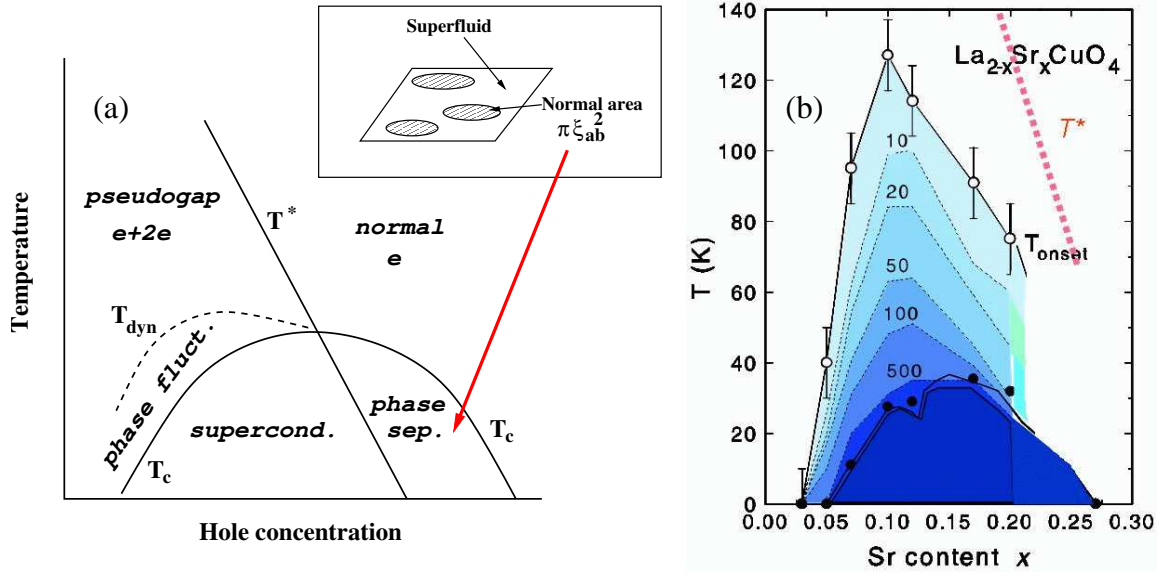}
\vskip 1.0 truecm

\label{Figure 8.} 

\caption{\label{Figure 8.}
(a) A generic phase diagram proposed for cuprates by 
Uemura [33], including the distinction between the 
pair formation $T^{*}$ and the onset temperature of 
dynamic superconductivity $T_{dyn}$ which corresponds to the 
$T_{onset}$ of the Nernst effect.
The inset illustrates the proposal of microscopic phase separation
between superconducting and normal metal regions in the 
overdoped region [79].
(b) The region of the Nernst effect, shown in the $T$-$x$ phase diagram
for LSCO [83].  The result of $T_{onset}$ for the $x=0.10$ sample
is plotted with the green star symbol in Fig. 2.  The $T^{*}$
values shown by the red-dashed line are taken from ref. [82].}
\end{center}
\end{figure}
\newpage
\begin{figure}[h]

\begin{center}

\includegraphics[angle=90,width=5.5in]{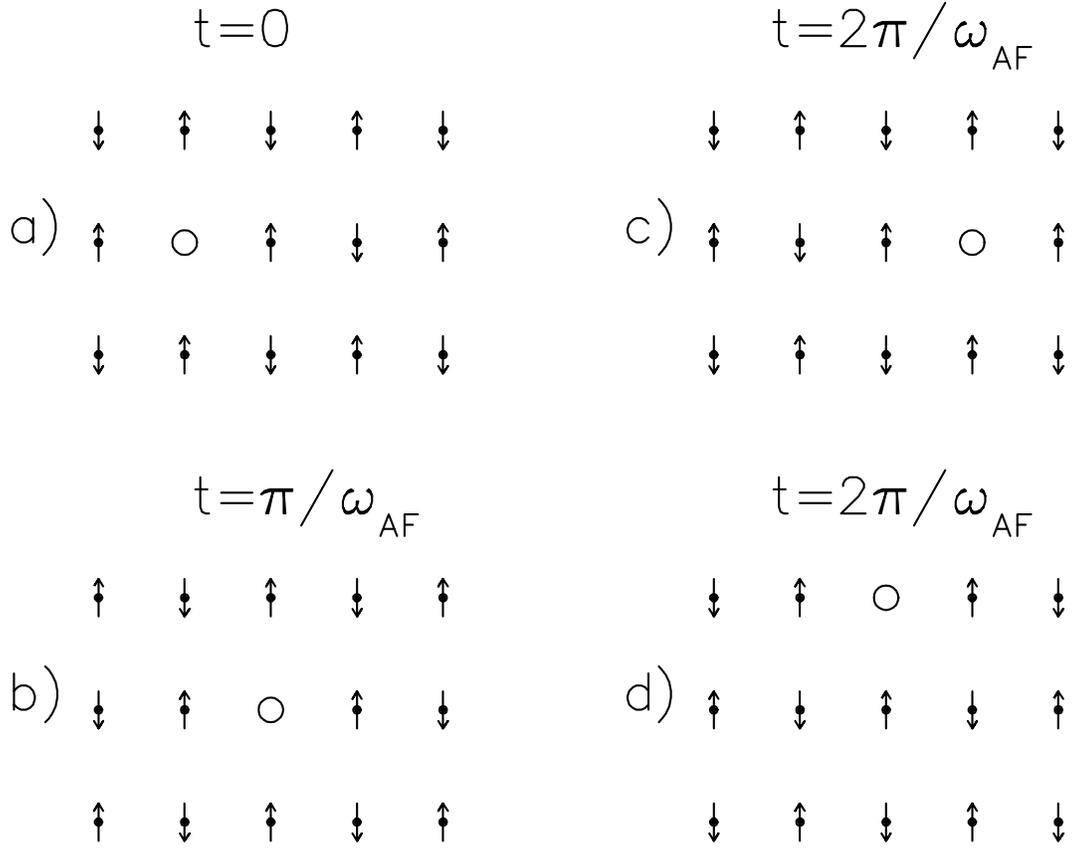}
\vskip 1.0 truecm

\label{Figure 9.} 

\caption{\label{Figure 9.}
An illustration of a charge hopping motion in the cuprate
resonant with the antiferromagnetic 
spin fluctuations with the frequency $\omega_{AF}$.
(a) is for the $t =0$ configuration, (b) 
for the $t=\pi/\omega_{AF}$ after a half period.
(c) and (d) show the situation after the full period
$t = 2\pi/\omega_{AF}$ for the propagation towards
the ($\pi,0$) direction and the ($\pi,\pi$) direction, respectively.
The charge motion sequenced with AF spin fluctuations helps avoid
extra frustration in the charge propagation process.}
\end{center}
\end{figure}
\newpage
\begin{figure}[h]

\begin{center}

\includegraphics[angle=0,width=5.5in]{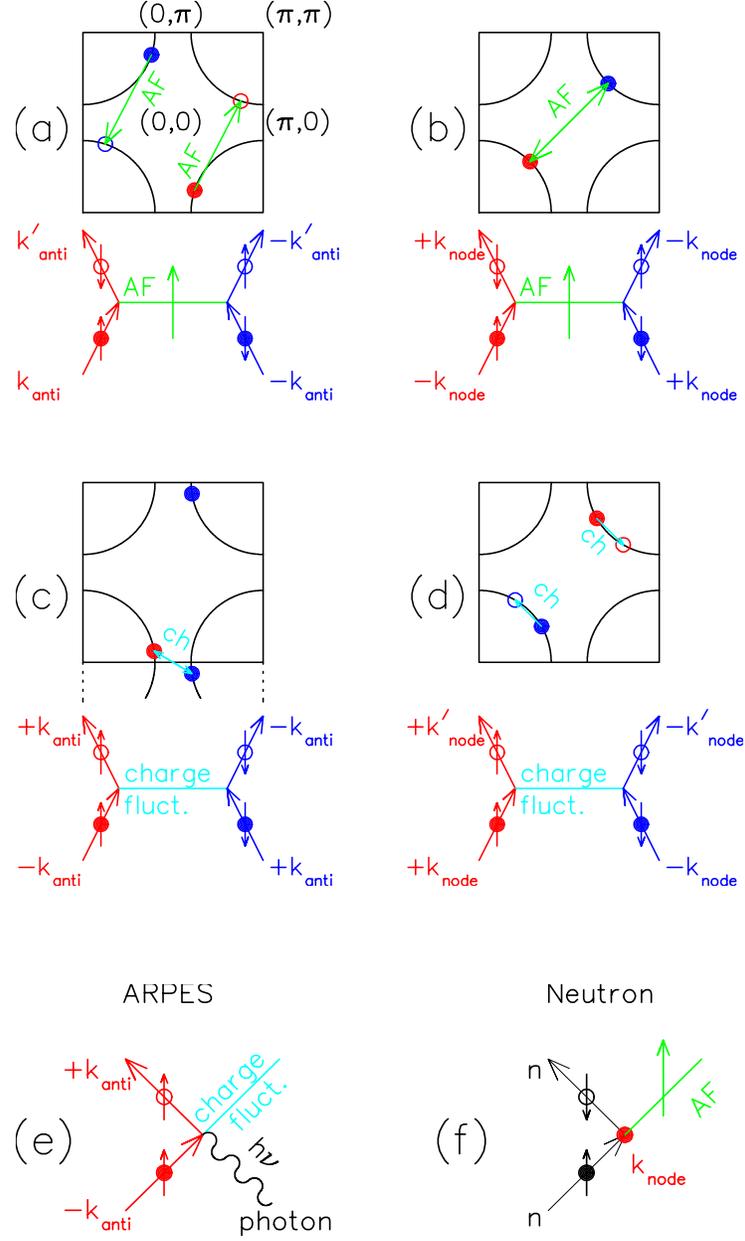}
\vskip 1.0 truecm

\label{Figure 10.} 

\caption{\label{Figure 10.}
An illustration of the attractive interaction 
obtained in the scattering process (a) and (d), and in the 
binding process (b) and (c) involving the exchange of 
AF spin fluctuations and charge fluctuations in the cuprates.
(e) and (f) show the diagrams for  
the ARPES coherence peak [23] for antinodal charges (e), and the 
neutron resonance
peak for nodal charges (f), which can be viewed as processes liberating
Yukawa-type bonding bosons.  These liberation processes 
occur at a cost of condensation energy,
corresponding to the hybrid spin/charge roton energy, with
the intensities proportional to the superfluid density $n_{s}/m^{*}$.
(a), (c) and (e) show processes for the antinodal charges, while
(b), (d) and (f) for nodal charges.}
\end{center}
\end{figure}

\end{document}